\documentclass[aps,prb,twocolumn,superscriptaddress,floatfix,amssymb,longbibliography]{revtex4-2}

\usepackage{graphicx}
\usepackage{dcolumn}
\usepackage{amsmath}
\usepackage{amssymb}
\usepackage{mathtools}

\usepackage[pdftex]{hyperref}
\hypersetup{colorlinks=true,linkcolor=blue,citecolor=blue,urlcolor=blue}
\usepackage{xcolor}
\newcommand{\black}{\color{black}}

\begin{document}

\title{${}^{31}$P NMR investigation of quasi-two-dimensional magnetic correlations in $T_2$P$_2$S$_6$~($T$~=~Mn~\&~Ni)}

\author{F. Bougamha} 
\affiliation{Department of Physics, Faculty of Sciences, University of Tunis El-Manar, Tunis 2092, Tunisia}
\affiliation{Institute for Solid State Research, Leibniz IFW Dresden, Helmholtzstr. 20, 01069 Dresden, Germany}

\author{S. Selter} 
\author{Y. Shemerliuk} 
\author{S. Aswartham} 
\affiliation{Institute for Solid State Research, Leibniz IFW Dresden, Helmholtzstr. 20, 01069 Dresden, Germany}

\author{A. Benali} 
\email{ali.benali@fst.utm.tn}
\affiliation{Department of Physics, Faculty of Sciences, University of Tunis El-Manar, Tunis 2092, Tunisia}

\author{B. B\"uchner} 
\affiliation{Institute for Solid State Research, Leibniz IFW Dresden, Helmholtzstr. 20, 01069 Dresden, Germany}
\affiliation{Institute of Solid State and Materials Physics and W\"{u}rzburg-Dresden Cluster of Excellence ct.qmat, Technische Universit\"{a}t Dresden, 01062 Dresden, Germany}

\author{H.-J. Grafe} 
\affiliation{Institute for Solid State Research, Leibniz IFW Dresden, Helmholtzstr. 20, 01069 Dresden, Germany}

\author{A. P. Dioguardi} 
\email{adioguardi@gmail.com}
\affiliation{Institute for Solid State Research, Leibniz IFW Dresden, Helmholtzstr. 20, 01069 Dresden, Germany}

\date{\today}

\begin{abstract}
We report the anomalous breakdown in the scaling of the microscopic magnetic susceptibility---as 
measured via the ${}^{31}$P nuclear magnetic resonance (NMR) shift $K$---with the bulk magnetic 
susceptibility $\chi$ in the paramagnetic state of Mn$_2$P$_2$S$_6$. This anomaly occurs near 
$T_\mathrm{max} \sim 117$\,K the maximum in $\chi(T)$ and is therefore associated with the onset of 
quasi-two-dimensional (quasi-2D) magnetic correlations. The spin--lattice relaxation rate divided by 
temperature $(T_1T)^{-1}$ in Mn$_2$P$_2$S$_6$ exhibits broad peak-like behavior as a function of 
temperature, qualitatively following $\chi$, but displaying no evidence of critical slowing down 
above the N\'{e}el temperature $T_N$.  In the magnetic state of Mn$_2$P$_2$S$_6$, NMR spectra provide 
good evidence for 60 degree rotation of stacking-fault-induced magnetic domains, as well as 
observation of the spin-flop transition that onsets at 4\,T. The temperature-dependent 
critical behavior of the internal hyperfine field at the P site in Mn$_2$P$_2$S$_6$ is consistent 
with previous measurements and the two-dimensional anisotropic Heisenberg model. In a sample of 
Ni$_2$P$_2$S$_6$, we observe only two magnetically split resonances in the magnetic state, 
demonstrating that the multiple-peaked NMR spectra previously associated with 60 degree rotation of 
stacking faults is sample dependent. Finally, we report the observation of a spin-flop-induced 
splitting of the NMR spectra in Ni$_2$P$_2$S$_6$, with an onset spin-flop field of 
$H_\mathrm{sf} = 14$\,T.
\end{abstract}

\maketitle

\section{Introduction}
\label{Introduction}

The $T_2$P$_2{Ch}_6$ family~\footnote{We denote the chemical formula using doubled notation due to the presence of the particular (P$_2{Ch}_6$)$^{4-}$ unit, which forms a structural motif in this class of compounds~\cite{Brec_1986_ReviewStructuralChemical, Piacentini_1984_SoftXray, Ouvrard_1988_Synthesiscrystalstructure}.} of quasi-two-dimensional (quasi-2D) magnetic van der Waals (vdWs) 
materials ($T$ = Mn, Ni, Fe, Co; and ${Ch}$ = S, Se) are currently the subject of broad 
and intense attention as a model system of low-dimensional magnetism, transport, and novel devices 
for technological applications in valleytronics~\cite{Li_2013_Couplingvalleydegree} and 
spintronics~\cite{Li_2019_IntrinsicVanDer}. These  magnetic vdWs materials have attracted 
significant interest, not only because of their technological significance, but also because they 
allow investigation of fundamental questions related to magnetism in low-dimensional 
systems~\cite{Burch_2018_Magnetismtwodimensional}. 

The $T_2$P$_2$S$_6$ subfamily (with $T$ = V, Mn, Fe, Co, Ni) has been investigated due to a variety 
of interesting physical and chemical properties, such as with regards to
anisotropy~\cite{Brec_1986_ReviewStructuralChemical, Grasso_1986_Opticalabsorptionspectra, 
Joy_1992_Magnetismlayeredtransition, Grasso_1990_Conductionprocesseslayered}, their potential 
applications as cathode materials for secondary batteries~\cite{Brec_1979_Physicalpropertieslithium}, 
ferroelectric properties~\cite{Simon_1994_ParaelectricFerroelectricTransition}, optically active nonlinear 
properties~\cite{Yitzchaik_1997_AnomalousSecondOrder,Lacroix_1994_StilbazoliumMPS_3Nanocomposites, 
Lagadic_1997_LayeredMPS_3M} and ion-exchange applications~\cite{Joy_1992_intercalationreactionpyridine}. 
Moreover, the ability of these quasi-2D vdWs materials to accommodate extrinsic intercalated 
species in the vdWs gap leads to a drastic change of their magnetic, electrical, and optical 
properties~\cite{Brec_1979_Physicalpropertieslithium}.

The first synthesis of a $T_2$P$_2$S$_6$ compound was reported by M. C. Friedel \textit{et al.} in 
1894~\cite{Friedel_1894_Surunenouvelle, Brec_1986_ReviewStructuralChemical}. But, only in the
1970’s, were more detailed structural studies carried 
out~\cite{Nitsche_1970_Crystalgrowthmetal, Taylor_1973_Preparationpropertiessome, Berthier_1978_NMRinvestigationlayered}. Recently, it has been discovered that 
Mn$_2$P$_2$S$_6$ and Ni$_2$P$_2$S$_6$ compounds are excellent platforms for investigating correlated 
electrons in 2D magnetic materials~\cite{Wang_2021_Spininducedlinear, 
Kim_2019_Suppressionmagneticordering, Kim_2018_ChargeSpinCorrelation}. Theoretical and experimental 
high-pressure research revealed the existence of insulator-to-metal transitions in 
Mn$_2$P$_2$S$_6$~\cite{Kim_2019_Suppressionmagneticordering, Wang_2016_PressureDrivenCooperative} 
and Ni$_2$P$_2$S$_6$ \cite{Kim_2019_Suppressionmagneticordering, Ma_2021_Dimensionalcrossovertuned}.
From a quasi-2D magnetic standpoint, Mn$_2$P$_2$S$_6$ has been studied in the antiferromagnetic (AFM) 
2D anisotropic Heisenberg model on a honeycomb lattice~\cite{Joy_1992_Magnetismlayeredtransition}. 
Whereas, theoretical investigations reveal that the magnetic anisotropy of Ni$_2$P$_2$S$_6$ is well 
treated by the XXZ model or a weakly anisotropic Heisenberg model~\cite{Joy_1992_intercalationreactionpyridine, 
Lancon_2016_Magneticstructuremagnon, Kim_2019_MottMetalInsulator}. 

At ambient pressure, Mn$_2$P$_2$S$_6$ is a highly resistive broad band semiconductor with a gap 
close to 3\,eV and is optically transparent and green in 
color~\cite{Grasso_2002_Lowdimensionalmaterials}. Mn$_2$P$_2$S$_6$ crystallizes in 
the monoclinic space group $C2/m$, with the Mn sites forming a honeycomb-like structure 
in the layer planes. The ambient temperature and pressure lattice parameters are $a = 6.05(1)$\,{\AA}, 
$b = 10.52(3)$\,{\AA}, $c = 6.80(2)$\,{\AA}, $\alpha = 90^{\circ}$ $\beta = 107.3(2)^{\circ}$, and 
$\gamma = 90^{\circ}$~\cite{Ressouche_2010_MagnetoelectricMnPS_3as}.

\begin{figure}[h!] 
    \includegraphics[trim=0cm 0cm 0cm 0cm, clip=true, width=0.9\linewidth]{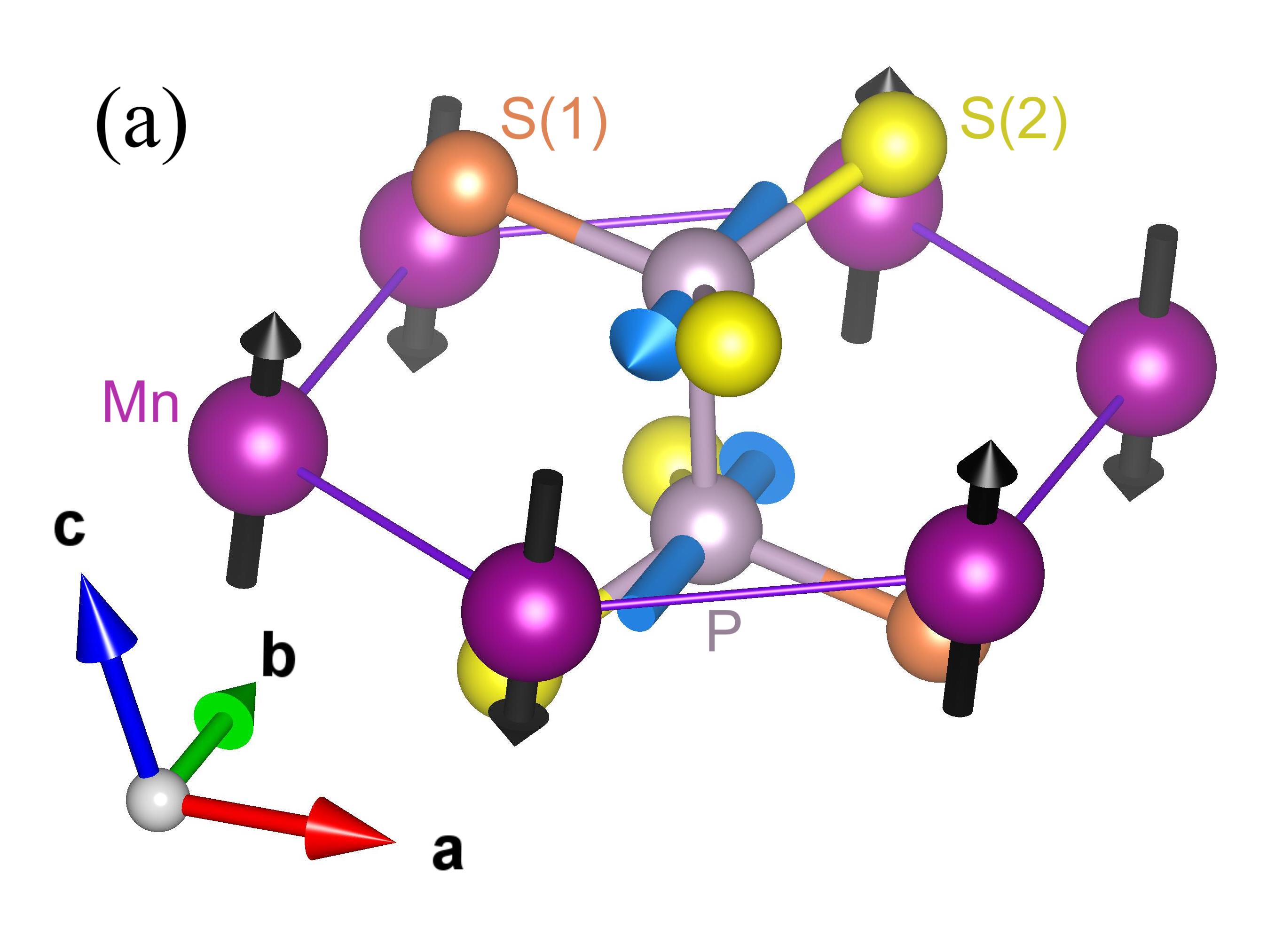}
    \includegraphics[trim=0cm 0cm 0cm 0cm, clip=true, width=0.9\linewidth]{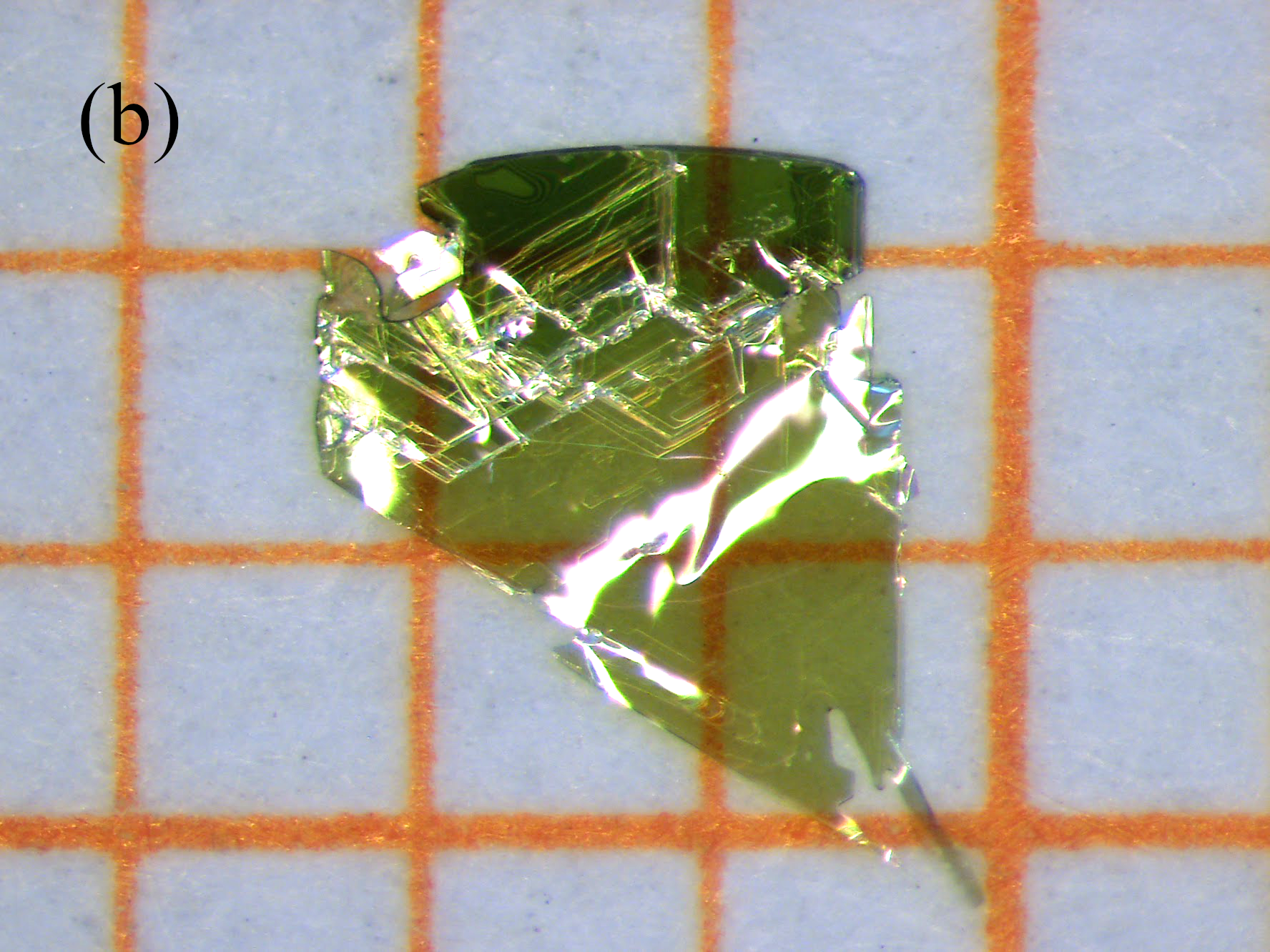}
    {\caption{\label{fig_1_Mn226_31P_local_environment_Crystal_B}(a) Local environment of the 
     magnetically inequivalent P sites in 
     Mn$_2$P$_2$S$_6$~\cite{Ouvrard_1985_Structuraldeterminationsome}. Mn are shown in purple, S(1) 
     in orange, S(2) in yellow, and P in pale pink. Black vectors represent the magnetic moments on 
     the Mn sites~\cite{Ressouche_2010_MagnetoelectricMnPS_3as}, and blue vectors represent the 
     calculated dipolar hyperfine field at the P sites due to the Mn magnetic 
     moments~\cite{Momma_2011_VESTA3three}. (b) Optical microscope image of Mn$_2$P$_2$S$_6$ 
     crystal B on a mm grid.}}
\end{figure}

Fig.\ref{fig_1_Mn226_31P_local_environment_Crystal_B}(a) shows local environment of the P dimers 
in the crystal structure. The weak interlayer coupling mediated by the S atoms, was proposed to be
purely of vdWs origin. The study of the transition from the paramagnetic state to the 
antiferromagnetic state collinear phase at 78\,K, on the other hand, indicates an interplane exchange 
related with some degree of metal-ligand covalency~\cite{Wildes_2007_Anisotropycriticalbehaviour, 
Wildes_2006_Staticdynamiccritical}. Neutron powder diffraction reveals a transition to the AFM state
 at 78\,K with the Mn moment of $4.43 \pm 0.03\,\mu_B$ canted $\sim 8^\circ$ from $c^*$ toward the $a$ direction~\cite{Ressouche_2010_MagnetoelectricMnPS_3as}, indicated by black vectors in Fig.~
\ref{fig_1_Mn226_31P_local_environment_Crystal_B}(a). $^{31}$P nuclear magnetic resonance (NMR) 
studies were previously carried out on several 
$T_2$P$_2$S$_6$~\cite{Berthier_1978_NMRinvestigationlayered, Ziolo_1988_31PNMRrelaxation,
Torre_1989_Spindynamicsmagnetic}, and revealed that NMR is a sensitive probe of magnetism. However, 
these studies were carried out on samples that showed significant magnetic field dependence of the
N\'{e}el temperature $T_N$.

Several recent studies have investigated the transition-metal-substitution dependence of the 
bulk magnetic properties of these compounds, revealing excellent tunability of the quasi-2D 
magnetic behavior in $T_2$P$_2$S$_6$ compounds~\cite{Selter_2021_Crystalgrowthanisotropic, 
Shemerliuk_2021_TuningMagneticTransport, Basnet_2021_Highlysensitivespin}. This tunability 
motivates comparison of our recent ${}^{31}$P nuclear magnetic resonance (NMR) study of
Ni$_2$P$_2$S$_6$~\cite{Dioguardi_2020_Quasitwodimensional} to Mn$_2$P$_2$S$_6$. Theoretical 
investigations should be focused on the relationship between the magnetic order and anisotropy 
in detail to the already experimentally investigated systems, such as 
(Mn$_{1-x}$Ni$_x$)$_2$P$_2$S$_6$ and 
(Mn$_{1-x}$Fe$_x$)$_2$P$_2$S$_6$~\cite{Masubuchi_2008_Phasediagrammagnetic, 
Shemerliuk_2021_TuningMagneticTransport}. An important piece of the puzzle is the so-called 
$K$--$\chi$ anomaly in Ni$_2$P$_2$S$_6$, which occurs in proximity to 
$T_\mathrm{max} \sim 262$\,K the maximum in the magnetic susceptibility, and indicates that 
the microscopic susceptibility observed via NMR provides unique local information about magnetic 
correlations. Systematic comparison of the interplay between magnetic anisotropy, magnetic 
interactions, and the $K$--$\chi$ anomaly in the quasi-2D magnets Ni and Mn end member compounds 
may help to shed light on the nature of the 
quasi-2d correlations in these systems.

In this work we investigate high-quality Mn$_2$P$_2$S$_6$ and Ni$_2$P$_2$S$_6$ single crystals 
that allow for sensitive ${}^{31}$P NMR measurements. We find that Mn$_2$P$_2$S$_6$ also exhibits a
$K$--$\chi$ anomaly, thus strengthening the conclusion that the anomaly is driven by quasi-2D 
magnetic correlations that emerge near the peak in the magnetic susceptibility 
($T_\mathrm{max} \sim 117$\,K) but above $T_N$. Furthermore, we observe effects of the quasi-2D 
magnetic fluctuations in Mn$_2$P$_2$S$_6$ via the nuclear spin--lattice relaxation rate divided by 
temperature $(T_1T)^{-1}$.  $(T_1T)^{-1}$ in qualitatively follows $\chi$, the peak-like shape of 
which is associated with quasi-two-dimensional magnetic correlations, indicating that $(T_1T)^{-1}$ 
is a good measure of quasi-2D fluctuations. However, similar in the case of Ni$_2$P$_2$S$_6$,
we do not observe any indication of critical slowing down above $T_N$. We also find that, in one 
of two measured crystals of Ni$_2$P$_2$S$_6$, there exist only two magnetically split peaks for all 
sample orientations. This indicates that the multiple-peaked NMR spectrum in the magnetic 
state---thought to be associated with 60 degree rotation of stacking faults---is sample dependent. 
Finally, we report the observation of a spin-flop transition in Ni$_2$P$_2$S$_6$ at 
$H_\mathrm{sf} = 14$\,T.

\section{Crystal Synthesis and Experimental Details}

\begin{figure} 
    \includegraphics[trim=0cm 0cm 0cm 0cm, clip=true, width=\linewidth]{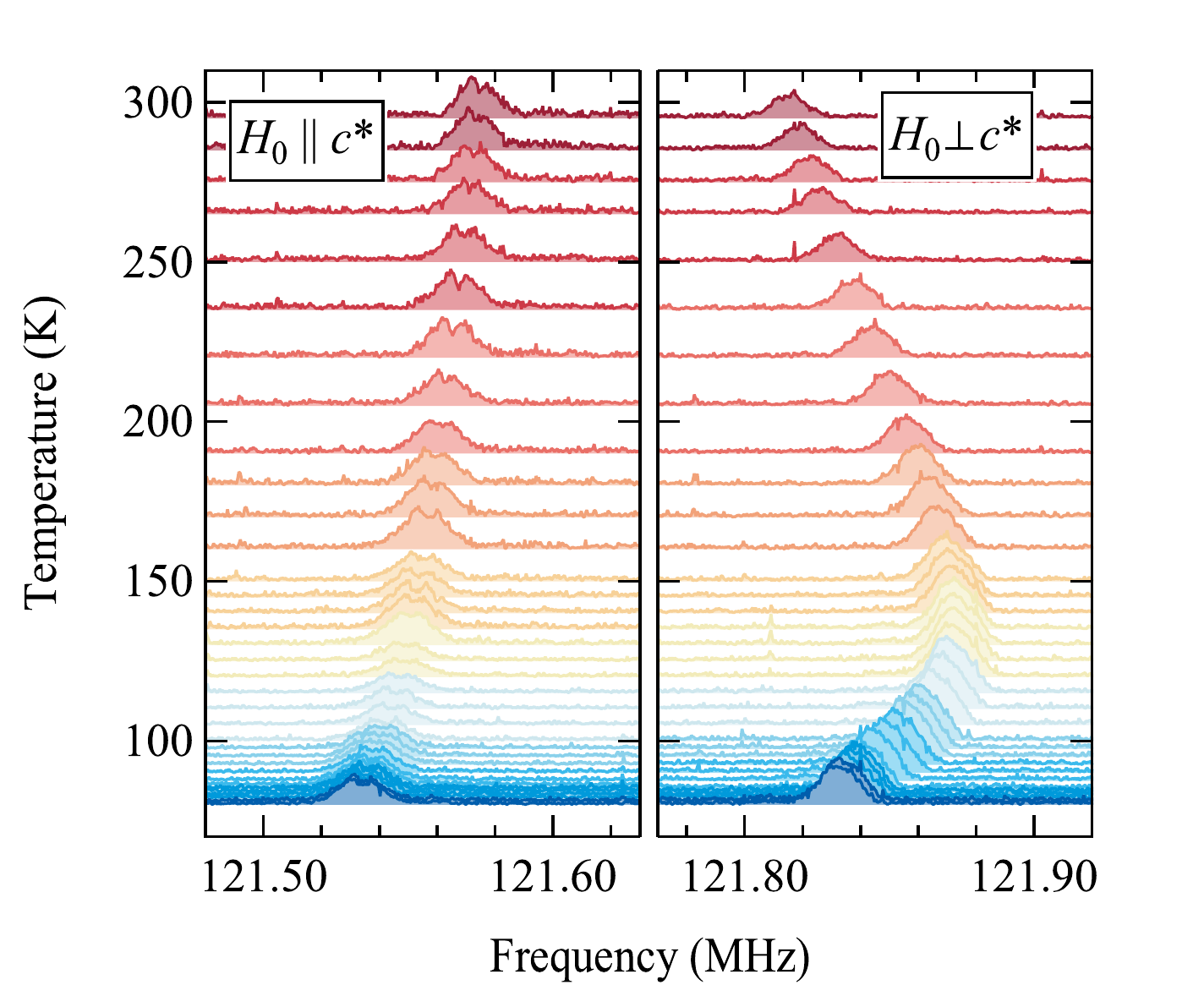}
    {\caption{\label{fig_2_Mn226_s2_spectra_waterfall}${}^{31}$P NMR frequency spectra as a 
     function of temperature from Mn$_2$P$_2$S$_6$ crystal B for (left) $H_0 \parallel c^*$ 
     and (right) $H_0 \perp c^*$ with nominal $H_0 = 7$\,T.}}
\end{figure}

Two crystals of Mn$_2$P$_2$S$_6$ were measured via NMR (hereafter referred to as Mn$_2$P$_2$S$_6$ 
crystals A and B. The crystals were plate-like, transparent and green in color, and on the order 
of 1\,mg (see Fig.~\ref{fig_1_Mn226_31P_local_environment_Crystal_B}(b)). One crystal of 
Ni$_2$P$_2$S$_6$ was remeasured in this work, and was previously described in 
Ref.~\cite{Dioguardi_2020_Quasitwodimensional}; we refer to this sample as Ni$_2$P$_2$S$_6$ 
crystal A, the same designation as the previous report. Details of the crystal growth via vapor 
transport and characterization can be found elsewhere~\cite{Shemerliuk_2021_TuningMagneticTransport, 
Dioguardi_2020_Quasitwodimensional, Selter_2021_Crystalgrowthanisotropic}. NMR was measured on the 
${}^{31}$P nuclei (spin $I = 1/2$, natural abundance 100\%, and gyromagnetic ratio 
${}^{31}\gamma/2\pi = 17.25144$\,MHz/T~\cite{Harris_2001_NMRnomenclatureNuclear}. All raw NMR shift 
values were calculated with respect to a ${}^{31}$P standard sample of 85\,\% H$_3$PO$_4$ in water 
($f_0 = 121.544 \pm 0.001$\,MHz). ${}^{33}$S and ${}^{55}$Mn NMR were also attempted, but without 
success. In the case of ${}^{33}$S this is likely due to the low natural abundance of the 
NMR-active isotope. In the case ${}^{55}$Mn, the large on-site magnetic moment is likely 
responsible for the lack of observability, which can result in a very large relaxation rate and/or 
shifts larger than 5\%. 

\begin{figure} 
    \includegraphics[trim=0cm 0cm 0cm 0cm, clip=true, width=\linewidth]{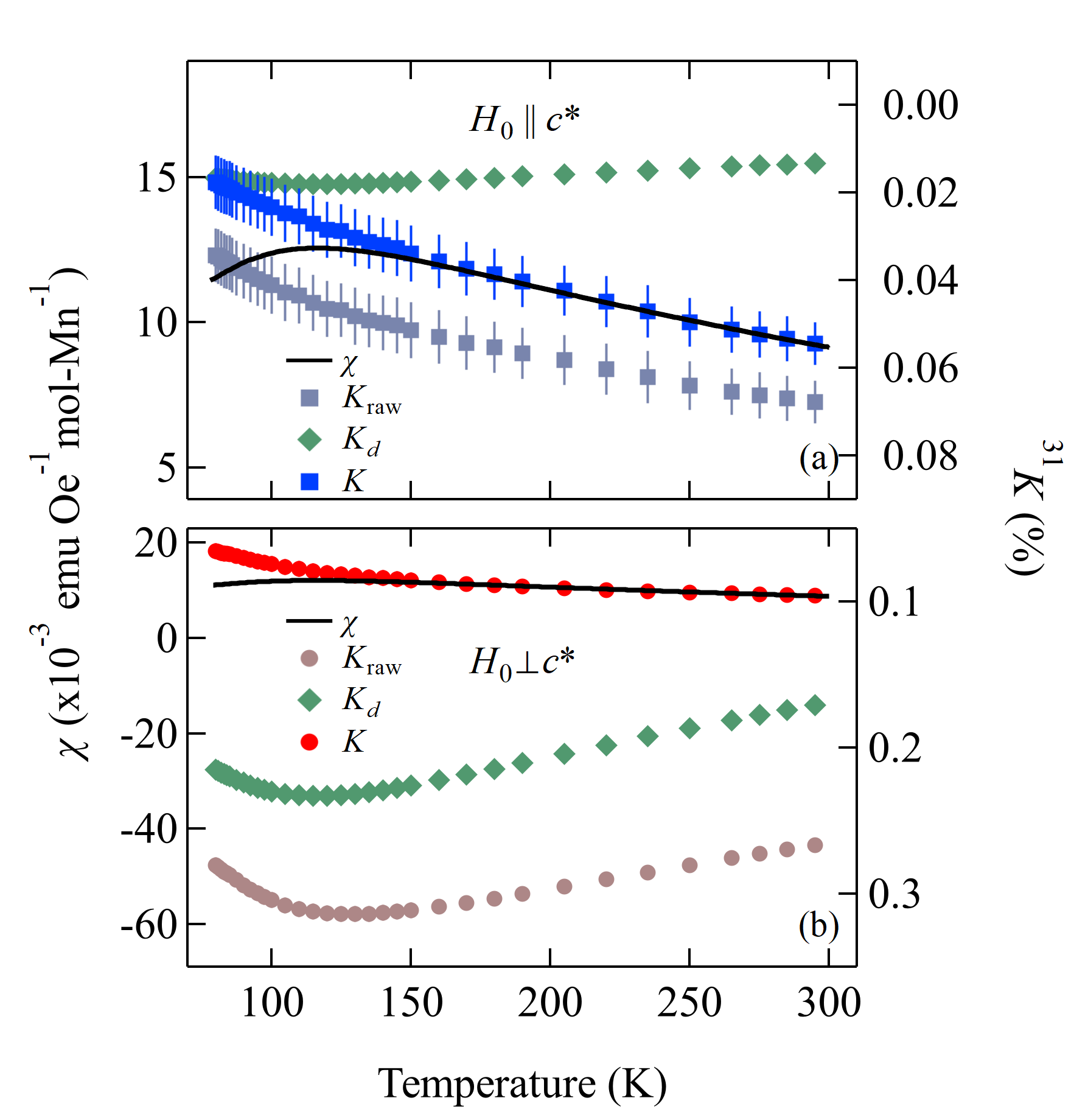}
    {\caption{\label{fig_3_Mn226_s2_K_and_chi_vs_temp}Mn$_2$P$_2$S$_6$ magnetic susceptibility 
     $\chi$ (solid lines) and NMR shift ${}^{31}K$ (markers) a function of temperature for (a) 
     $H_0 \parallel c^*$ and (b) $H_0 \perp c^*$. The axes have been scaled to highlight the 
     deviation from scaling below approximately 150\,K and clearly outside of the experimental 
     uncertainties below $T_\mathrm{max}$. The raw shift $K_\mathrm{raw}$ was extracted from 
     Gaussian fits to the data in Fig.~\ref{fig_2_Mn226_s2_spectra_waterfall}, with respect the 
     ${}^{31}$P in 85\,\% H$_3$PO$_4$ in water. The shift $K = K_\mathrm{raw} - K_d$, where $K_d$ 
     is the contribution from bulk magnetic effects as discussed in the text.}}
\end{figure}

We note that Mn$_2$P$_2$S$_6$ crystal A was slightly deformed upon insertion into the NMR coil. 
This resulted in broadening of the spectra into a double-peak in the paramagnetic state. This 
double peak was absent in Mn$_2$P$_2$S$_6$ crystal B, which was pristine. While measurements in 
the magnetic state were unaffected by this physical deformation, the paramagnetic state 
measurements required a double-peak Gaussian fit to characterize the spectra. Therefore, all 
normal-state spectral measurements shown in this work were conducted on Mn$_2$P$_2$S$_6$ crystal B. 
We note however, that double-peak fits to the spectra from Mn$_2$P$_2$S$_6$ crystal A yield 
identical hyperfine couplings, temperature dependence, and angular dependence.

NMR spectra were collected via standard spin-echo and free-induction-decay (FID) pulse sequences as
function of temperature, frequency, and angle. In the magnetic state, when the spectrum becomes too 
broad to acquire with a single spin echo or FID, we utilize computer-controlled stepper motors to 
tune and match the resonant circuit and sum the resulting Fourier transforms to reconstruct the 
intrinsic spectra.

\subsection{Paramagnetic State Measurements}

\begin{figure} 
    \includegraphics[trim=0cm 0cm 0cm 0cm, clip=true, width=\linewidth]{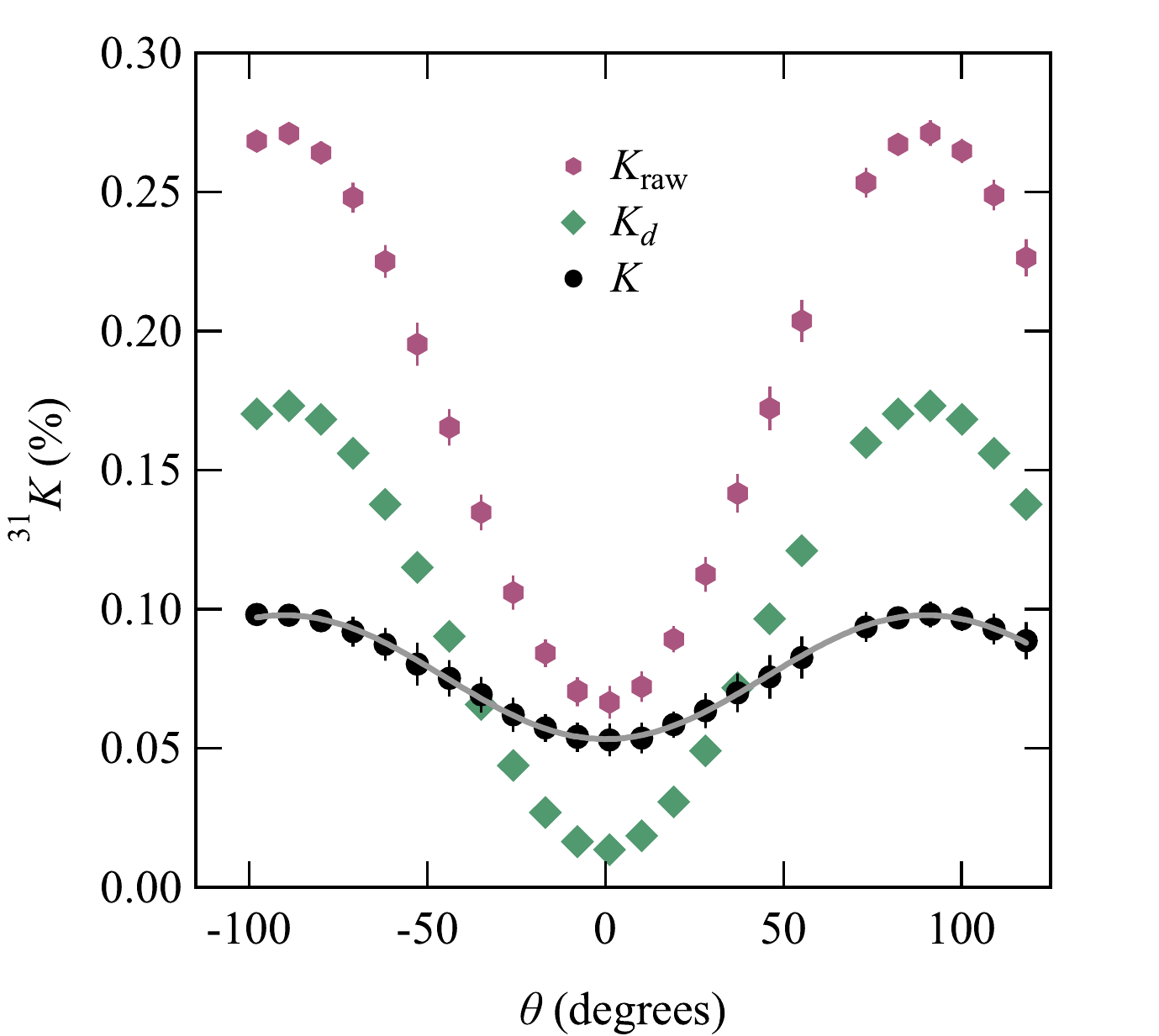}
    {\caption{\label{fig_4_Mn226_s2_288K_K_ang_dep}Mn$_2$P$_2$S$_6$ NMR shift ${}^{31}K$ as a 
     function of out-of-plane angle $\theta$ (black circles), where  $\theta = 0^\circ$ indicates 
     $H_0 \parallel c^*$. The grey curve is a fit to Eqn.~\ref{K_vs_theta_eqn} as described in the 
     text, which yields the fit parameters $K_{c^*} = 0.053 \pm 0.002$\,\% and 
     $K_{ab} = 0.098 \pm 0.001$\,\%. The raw shift $K_\mathrm{raw}$ (salmon hexagons) and the
     contribution from bulk magnetic effects $K_d$ (green diamonds) are shown for completeness.}}
\end{figure}

\begin{figure*} 
    \includegraphics[trim=0cm 0cm 0cm 0cm, clip=true, width=\linewidth]{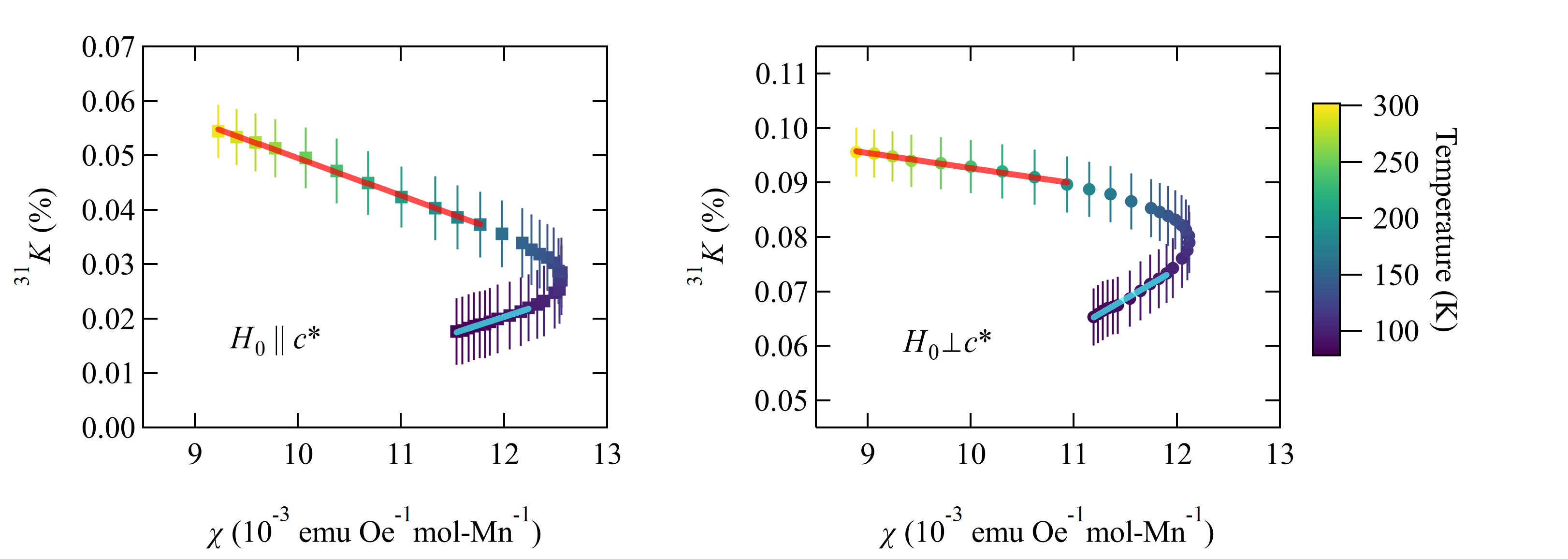}
    {\caption{\label{fig_5_Mn226_s2_K_vs_chi}Mn$_2$P$_2$S$_6$ NMR shift ${}^{31}K$ as a function 
     of magnetic susceptibility $\chi$ with temperature as an implicit parameter. Lines are fits 
     to extract the hyperfine coupling $A$ and orbital shift $K_0$ as summarized in Table~\ref{hyperfine_fit_pars_table}.}}
\end{figure*}

NMR spectra as a function of temperature are shown in Fig.~\ref{fig_2_Mn226_s2_spectra_waterfall} 
for both $H_0 \parallel c^*$ and $H_0 \perp c^*$. The raw NMR shifts $K_\mathrm{raw}$ were 
extracted with respect H$_3$PO$_4$, and bulk magnetic effects $K_d$ were subtracted, assuming an 
ellipsoidal crystal shape as described previously~\cite{Dioguardi_2020_Quasitwodimensional,Osborn_1945} using 
the approximate crystal dimensions of Mn$_2$P$_2$S$_6$ crystal B ($L_x = 2.2$\,mm, $L_y = 1.6$\,mm, 
and $L_z = 0.1$\,mm). The resulting NMR shifts $K$ are shown in 
Fig.~\ref{fig_3_Mn226_s2_K_and_chi_vs_temp} as a function of temperature along with the bulk 
magnetic susceptibility (measured in the same applied field of 7\,T). The high-temperature data 
have been scaled to show the striking $K$--$\chi$ anomaly that manifests below 150\,K.

We also measured the angular dependence of the NMR spectrum at 295\,K, and perform the same 
correction for bulk magnetization effects, with angular dependence of the susceptibility included. 
The results are shown in Fig.~\ref{fig_4_Mn226_s2_288K_K_ang_dep}. There is no in-plane angular 
dependence to within the experimental error, therefore, considering also the axial symmetry of the 
local P environment, we make the assumption that the NMR shift tensor is axially symmetric, i.e. 
$K_a = K_b = K_{ab} \neq K_{c^*}$ . The grey curve in Fig.~\ref{fig_4_Mn226_s2_288K_K_ang_dep} 
is a least-squares fit to the expected out-of-plane angular dependence
\begin{equation}
\label{K_vs_theta_eqn}
K(\theta) = K_\mathrm{iso} + K_\mathrm{ax}\left(3\cos^2{\left[(\theta - \theta_0)\frac{\pi}{180}\right]} - 1\right),
\end{equation}
where $K_\mathrm{iso} = \frac{1}{3}(2K_{ab} + K_{c^*})$ and $K_\mathrm{ax} = \frac{1}{3}(K_{c^*} - K_{ab})$. 
The resulting best fit is described by the parameters $K_{c^*} = 0.053 \pm 0.002$\,\% and 
$K_{ab} = 0.098 \pm 0.001$\,\%.

To better highlight the shift anomaly, we also plot $K$ as a function of $\chi$ with temperature as 
an implicit parameter in Fig.~\ref{fig_5_Mn226_s2_K_vs_chi}. One expects $K \propto \chi$, and 
therefore linear relations on such plots (also known as Clogston-Jaccarino plots~
\cite{Clogston_1964_InterpretationKnightShifts}). The constant----which in the case of an anisotropic material, is actually a tensor quantity---of proportionality in an uncorrelated material is the hyperfine coupling $\mathbf{A}$. In the case of Mn$_2$P$_2$S$_6$, however, we observe two different regions of
linear behavior with a change in the sign of the slope in between. The linear fits shown in red and
blue in Fig.~\ref{fig_5_Mn226_s2_K_vs_chi} are characterized by the fit 
parameters in Table~\ref{hyperfine_fit_pars_table}. The fit ranges are as follows: for 
$H_0 \parallel c^*$ the high temperature fit range is $175\,\mathrm{K} \leq T \leq 285\,\mathrm{K}$ 
and the low temperature range is $80\,\mathrm{K} \leq T \leq 93\,\mathrm{K}$, and for $H_0 \perp c^*$ 
the high temperature fit range is $175\,\mathrm{K} \leq T \leq 295\,\mathrm{K}$ and the low 
temperature range is $80\,\mathrm{K} \leq T \leq 102\,\mathrm{K}$. We note that the low temperature 
fits produce a ``hyperfine coupling,'' but we hesitate to ascribe meaning to this, as will be 
discussed below.

\begin{table}
    \begin{ruledtabular}
    \begin{tabular}{ccc}
             & High Temperature                 & Low Temperature                    \\[0.25ex]
    \hline \\[-1.5ex]
    $A_{c^*}$    & $-0.04 \pm 0.01\,\mathrm{T}/\mu_B$ & $ 0.03 \pm 0.05\,\mathrm{T}/\mu_B$ \\[0.5ex]
    $A_{ab}$     & $-0.02 \pm 0.01\,\mathrm{T}/\mu_B$ & $ 0.06 \pm 0.04\,\mathrm{T}/\mu_B$ \\[0.5ex]
    $K_{0, c^*}$ & $ 0.12 \pm 0.02\,\%$               & $-0.1 \pm 0.1\,\%$               \\[0.5ex]
    $K_{0, ab}$  & $ 0.12 \pm 0.02\,\%$               & $-0.06 \pm 0.08\,\%$               \\[0.5ex]
    \end{tabular}
    \end{ruledtabular}
    {\caption{\label{hyperfine_fit_pars_table}Mn$_2$P$_2$S$_6$ hyperfine coupling constants $A$ and 
     orbital shifts $K_0$ extracted from fits to ${}^{31}K$ vs $\chi$ as shown in 
     Fig~\ref{fig_5_Mn226_s2_K_vs_chi}.}}
\end{table}

\begin{figure}
    \includegraphics[trim=0cm 0cm 0cm 0cm, clip=true, width=\linewidth]{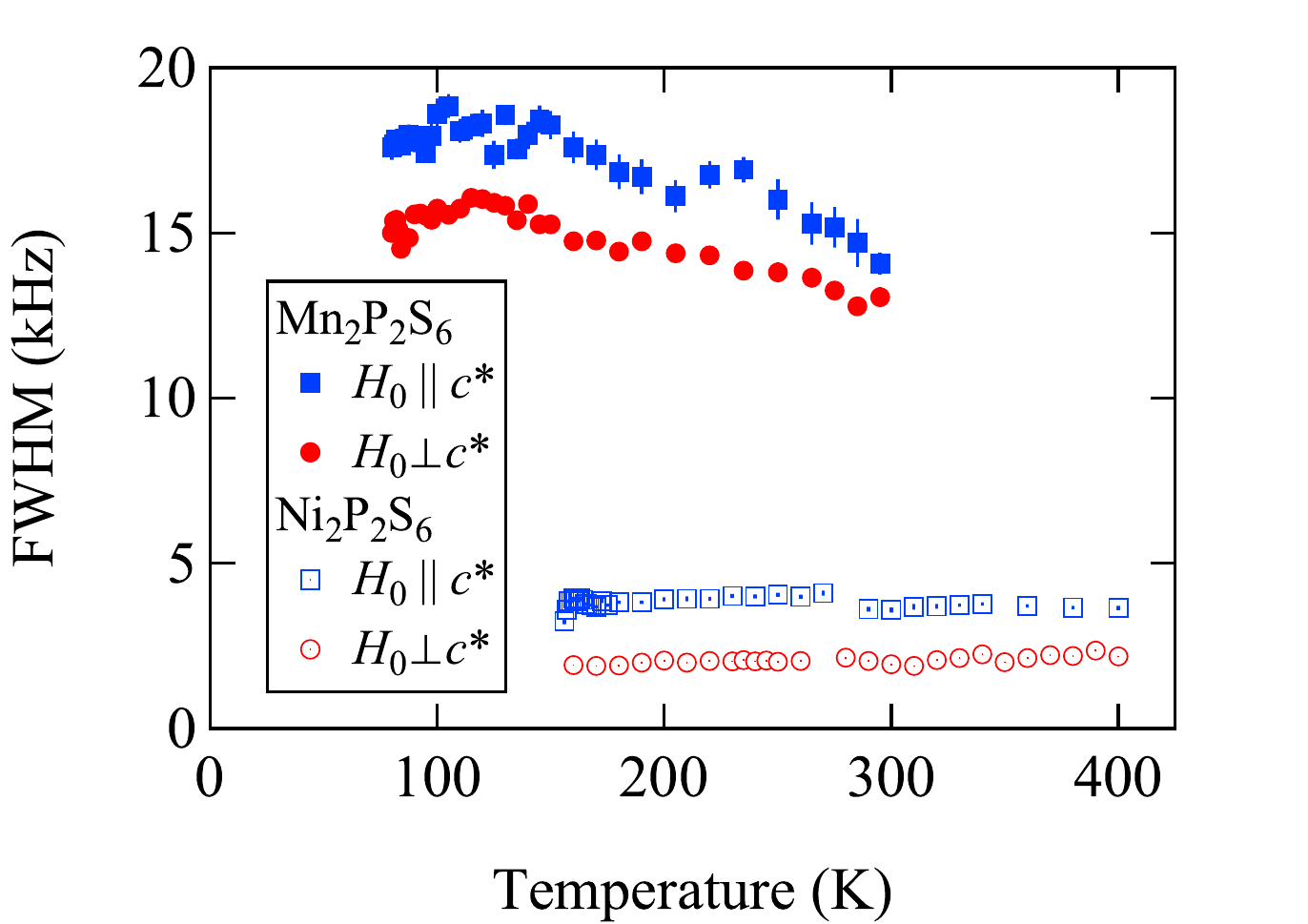} 
    {\caption{\label{fig_6_MnNi226_FWHM_vs_temp}Full width at half maximum (FWHM) of the 
     ${}^{31}$P resonances for Mn$_2$P$_2$S$_6$ and Ni$_2$P$_2$S$_6$ as a function of temperature 
     in the paramagnetic state. Ni$_2$P$_2$S$_6$ data are from 
     Ref.~\cite{Dioguardi_2020_Quasitwodimensional} Mn$_2$P$_2$S$_6$ data are from single-peak 
     fits to the spectra in Fig.~\ref{fig_2_Mn226_s2_spectra_waterfall}.}}
\end{figure}

As mentioned above, we find that the full width at half maximum (FWHM) of the spectra in 
Mn$_2$P$_2$S$_6$ is significantly larger than in Ni$_2$P$_2$S$_6$ (see 
Fig.~\ref{fig_6_MnNi226_FWHM_vs_temp}). Considering the large magnitude of the bulk magnetic 
susceptibility, and the FWHM scales qualitatively with the bulk magnetic susceptibility, we 
conclude that this is the dominant contribution to the increased linewidth. 

\subsection{Antiferromagnetic State Measurements}

\begin{figure*}  
    \includegraphics[trim=0cm 0cm 0cm 0cm, clip=true, width=\linewidth]{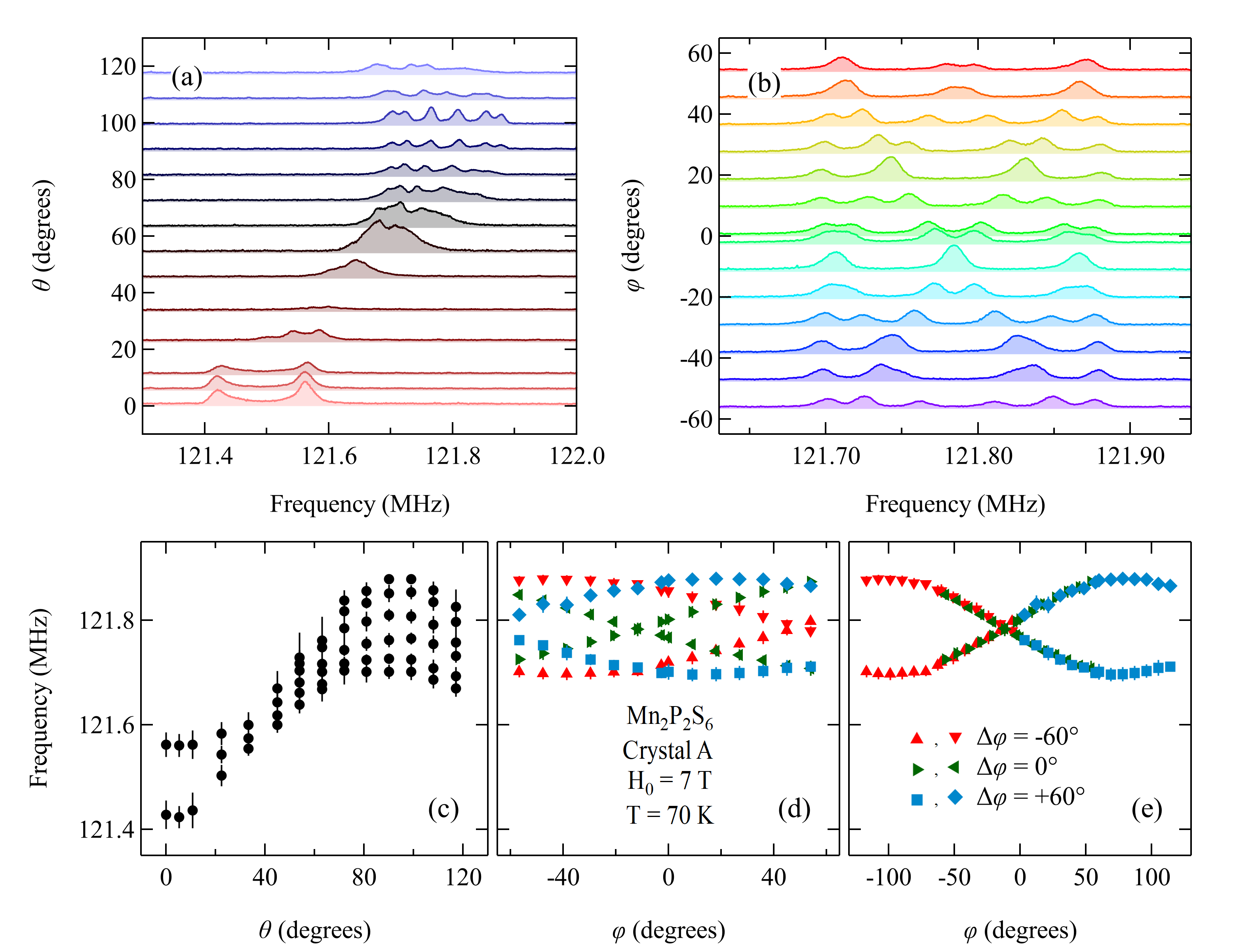}
    {\caption{\label{fig_7_Mn226_s1_magnetic_state_spectra}Mn$_2$P$_2$S$_6$ ${}^{31}$P NMR spectra 
     offset by out-of-plane angle $\theta$ ($\theta = 0$\,degrees corresponds to $H_0 \parallel c^*$) (a) and in-plane angle $\varphi$ in the magnetic state 
     at $T = 70$\,K. (c) Frequencies vs $\theta$ extracted from multi-peak fits to the spectra in 
     (a). (d) Frequencies vs $\varphi$ extracted from (b) by the same method. (e) In-plane rotation 
     center frequencies from (d) offset by $-60$, $0$, and $60$ degrees. The markers/colors in (d) 
     and (e) were chosen by hand to indicate the unique resonances.}}
\end{figure*}

\begin{figure}  
    \includegraphics[trim=0cm 19cm 8.5cm 0cm, clip=true, width=\linewidth]{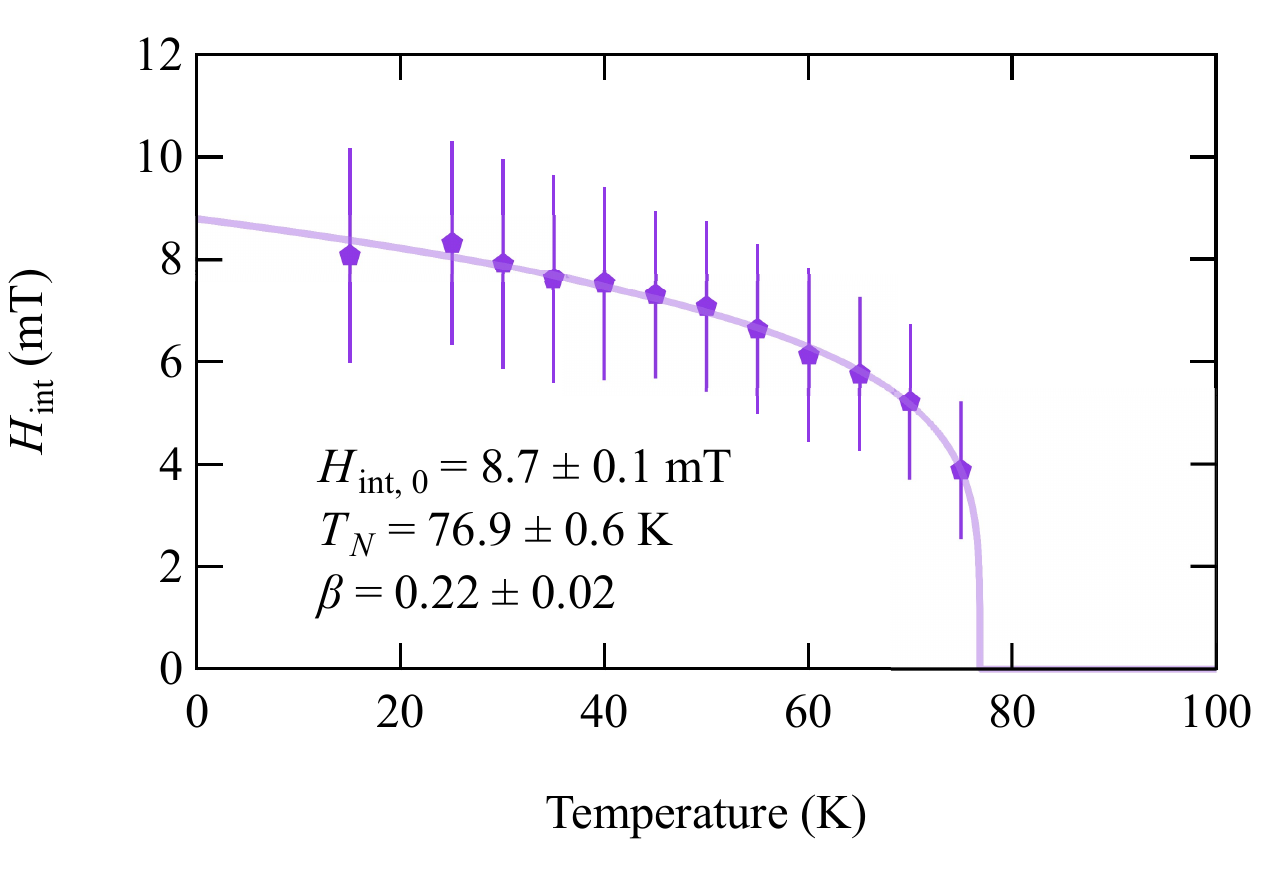}
    {\caption{\label{fig_8_Mn226_Hint_vs_temp}Temperature dependence of the internal field 
     $H_{\mathrm{int}}$ obtained from NMR spectra of Mn$_2$P$_2$S$_6$ measured for $H_0\perp c^*$ 
     in the magnetic state. The light purple line is the fit to 
     Equation~\ref{internal_field_fit_function} as described in the text. Error bars indicate 
     $\sigma$ of the distribution of internal fields, uncertainties of peak value of 
     $H_{\mathrm{int}}$ are smaller than the marker size.}}
\end{figure}

NMR is an excellent probe of magnetic order; the key measureable parameter is the internal 
hyperfine field at the P site. We measured the ${}^{31}$P NMR spectra as a function of angle and 
temperature in the antiferromagnetic state. The angular dependence of the spectra is show in 
Fig.~\ref{fig_7_Mn226_s1_magnetic_state_spectra}. We perform multipeak fitting to extract the 
angular dependence of the resonances. The spectra were collected in an external field of 7\,T, 
which is above the spin-flop field $T_{\mathrm{sf}} = 4$\,T. As a result, the spectra collapse 
to two split peaks for $H_0$ nearly parallel to the $c^*$ direction as the internal hyperfine 
field is perturbed by the 7\,T external field's influence on the Mn moments. The in-plane rotation
spectra are not significantly perturbed, as the external field aligned in the basal plane is
approximately perpendicular to the easy axis, and therefore does not result in a spin flop. We 
find six magnetically split peaks, that fall onto a single curve when offset by $\pm 60^\circ$. 
This phenomenon was also observed in the crystal B of 
Ni$_2$P$_2$S$_6$~\cite{Dioguardi_2020_Quasitwodimensional}, and has been associated with 
stacking faults. We also tracked the internal hyperfine field as a function of temperature in the
magnetic state, as shown in Fig.~\ref{fig_8_Mn226_Hint_vs_temp}.

The internal field allows us to use NMR to carefully measure the spin-flop transition as a 
function of applied external field. We first aligned the Mn$_2$P$_2$S$_6$ in the normal state 
such that the applied external field was nearly parallel to $c^*$, and then cooled the crystal 
to 65.3\,K. We fit the two resulting magnetically split peaks to a double Gaussian function, 
holding the amplitudes and widths equal. We extract the component of the internal field along 
the $c^*$ direction, which is approximately zero at low fields, but begins to grow starting 
at $H_{\mathrm{sf}} = 4$\,T and saturates above 7\,T. We note that previous measurements find 
equivalent field dependence associated with a spin-flop transition at low temperatures, 
indicating that our NMR measurements at higher temperatures in the AFM state are 
valid~\cite{Okuda_1986_MagneticPropertiesLayered, Goossens_2000_Orderingnaturespin}. We choose 
to measure at high temperature due to the extremely slow spin--lattice relaxation rates deep in 
the AFM state. 

\begin{figure} 
    \includegraphics[trim=0cm 0cm 0cm 0cm, clip=true, width=\linewidth]{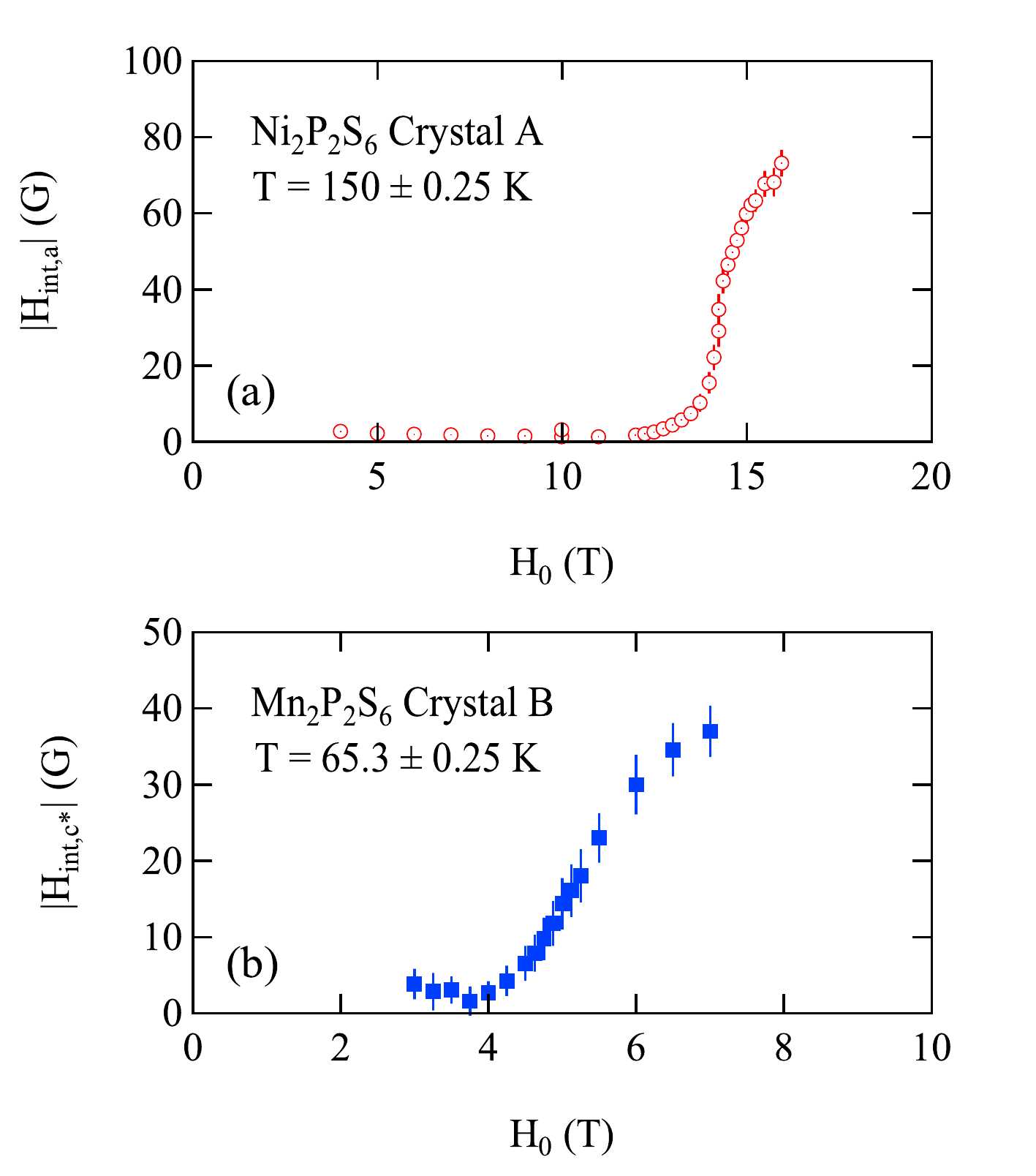}
    {\caption{\label{fig_9_MnNi226_spin_flop}Component of the internal field parallel 
     to the applied external field as a function of $H_0$, showing evidence of the spin-flop 
     transition, which begins at (a) 14\,T in the case of Ni$_2$P$_2$S$_6$ with $H_0$ aligned 
     approximately along $a$ and (b) 4\,T in the case of Mn$_2$P$_2$S$_6$ for $H_0$ aligned along 
     approximately with $c^*$.}}
\end{figure}

\begin{figure} 
    \includegraphics[trim=0cm 0cm 0cm 0cm, clip=true, width=\linewidth]{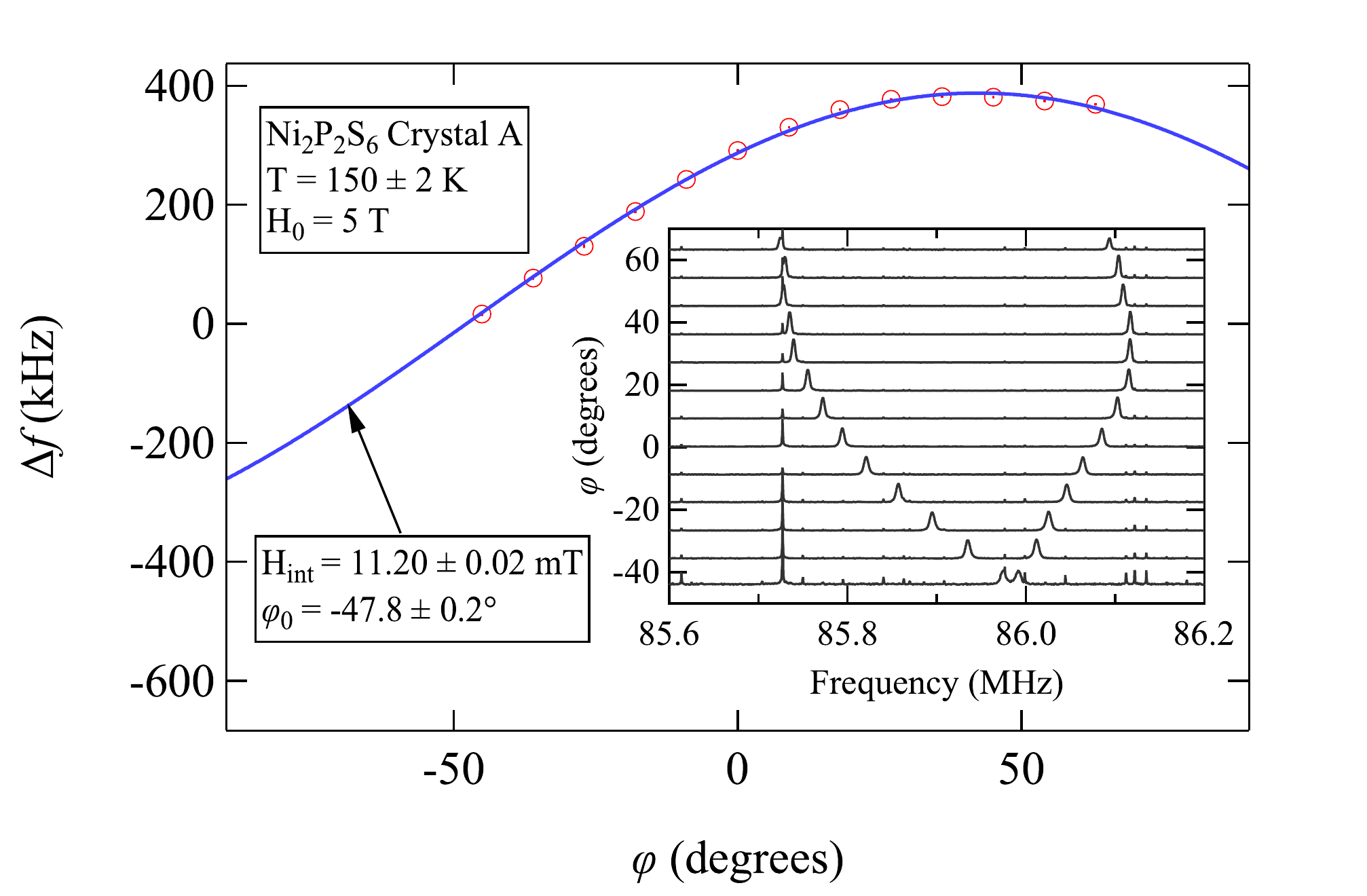}
    {\caption{\label{fig_10_Ni226_s1_150K_in-plane_ang_dep}In-plane angular dependence of the 
     ${}^{31}$P magnetic peak splitting $\Delta f$ as a function of in-plane angle $\varphi$ for 
     Ni$_2$P$_2$S$_6$ crystal A at $T = 150 \pm 2$\,K and $H_0 = 5$\,T. The inset shows the raw 
     frequency-swept spectra. Note that the magnet was in persistent mode for these measurements, 
     and the field likely drifted enough to shift the spectra from one measurement to the next. 
     This has no effect on the splitting. Hint = 10.5 mT for sample B.}}
\end{figure}

We then conducted the same experiment on Ni$_2$P$_2$S$_6$ crystal A, but with $H_0$ along $a$,
which remarkably yield a similar result, summarized in Fig.~\ref{fig_9_MnNi226_spin_flop}. 
These results provide good evidence for a previously unreported spin-flop transition in 
Ni$_2$P$_2$S$_6$ with a spin-flop field of $H_{\mathrm{sf}} = 14$\,T. Furthermore, in 
Ni$_2$P$_2$S$_6$ crystal A, we also discovered that there are only two magnetically split 
peaks at 5\,T for all in-plane crystal orientations (see 
Fig.~\ref{fig_10_Ni226_s1_150K_in-plane_ang_dep}), indicating that stacking faults are not 
present in this crystal. Therefore, we conclude that the presence of stacking faults is sample 
dependent. The spin-flop transition exists in both Mn$_2$P$_2$S$_6$ and Ni$_2$P$_2$S$_6$ in 
different field orientations, which is associated with their magnetic moment direction. As shown 
in the Fig.~\ref{fig_1_Mn226_31P_local_environment_Crystal_B}(a) and Fig.~1 in 
Ref.~\cite{Dioguardi_2020_Quasitwodimensional}, respectively. The Mn moments in Mn$_2$P$_2$S$_6$ 
are oriented in a mostly out-of-plane orientation, while the Ni moments in Ni$_2$P$_2$S$_6$ are oriented 
mostly in the $a$-direction, with a small component along the $c$ 
direction~\cite{Wildes_2015_Magneticstructurequasi}. In collinear AFM systems, a magnetic field 
along the easy axis that exceeds a critical spin-flop field $H_{\mathrm{sf}}$ induces the magnetic 
moments to rotate~\cite{Basnet_2021_Highlysensitivespin}.

\subsection{Relaxation Measurements}

\begin{figure}[h!] 
    \includegraphics[trim=0cm 0cm 0cm 0cm, clip=true, width=\linewidth]{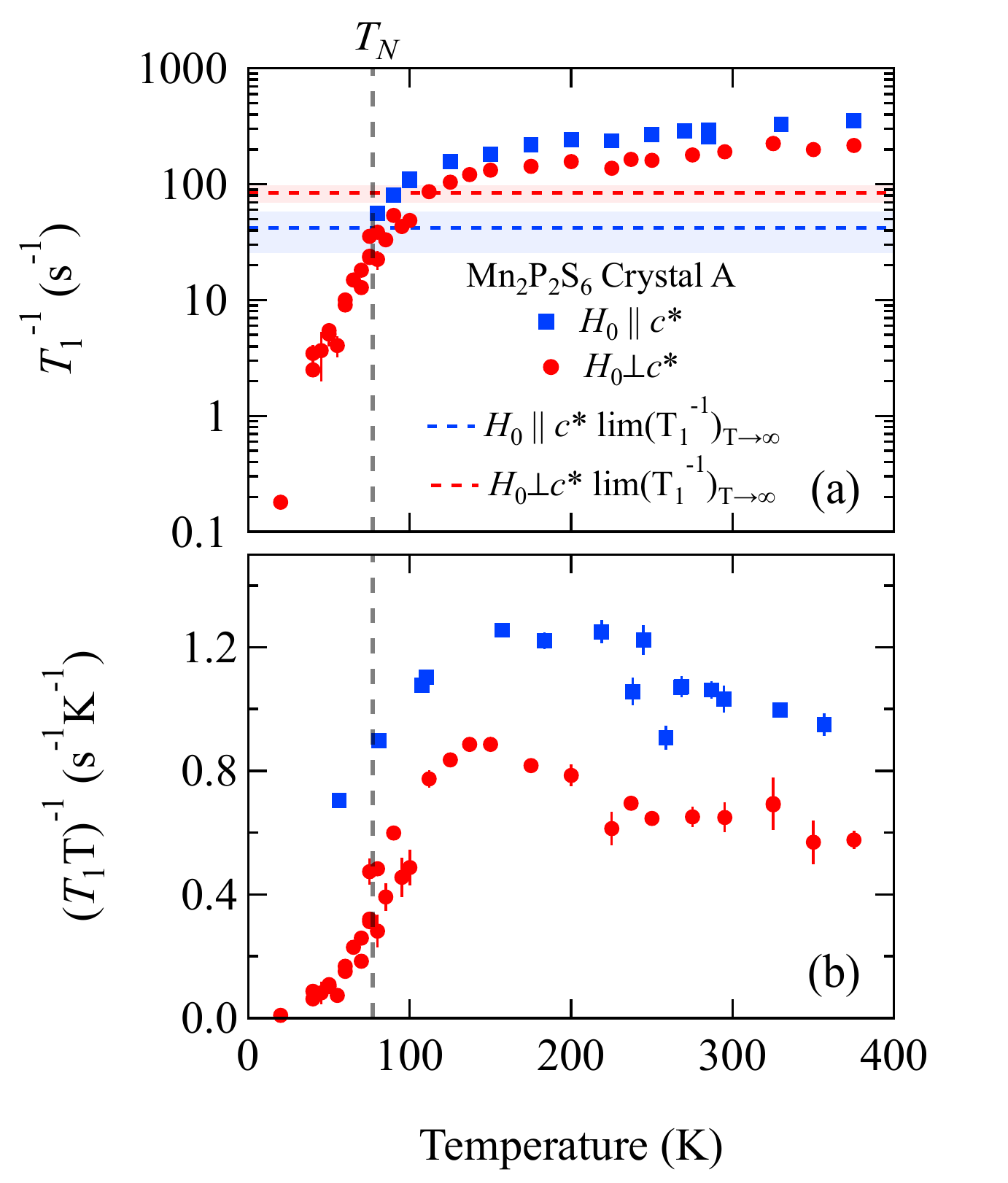} 
    {\caption{\label{fig_11_Mn226_T1}(a) $(T_1)^{-1}$ vs temperature from  Mn$_2$P$_2$S$_6$
     crystal A. (b) Spin--lattice relaxation rate divided by temperature $(T_1T)^{-1}$ as 
     a function of temperature for the same sample. The Néel temperature $T_{N}$ is 
     marked with a dashed vertical line. }}
\end{figure}

The spin--lattice relaxation rate was measured for both orientations of Mn$_2$P$_2$S$_6$ 
crystal A as a function of temperature via saturation recovery, and the data are summarized 
in Fig.~\ref{fig_11_Mn226_T1}. The spin--lattice relaxation rate divided by 
temperature $(T_1T)^{-1}$ increases slightly with decreasing temperature down to approximately 
150\,K, and then decreases, qualitatively following the bulk magnetic susceptibility.

\begin{figure}
    \includegraphics[width=\linewidth]{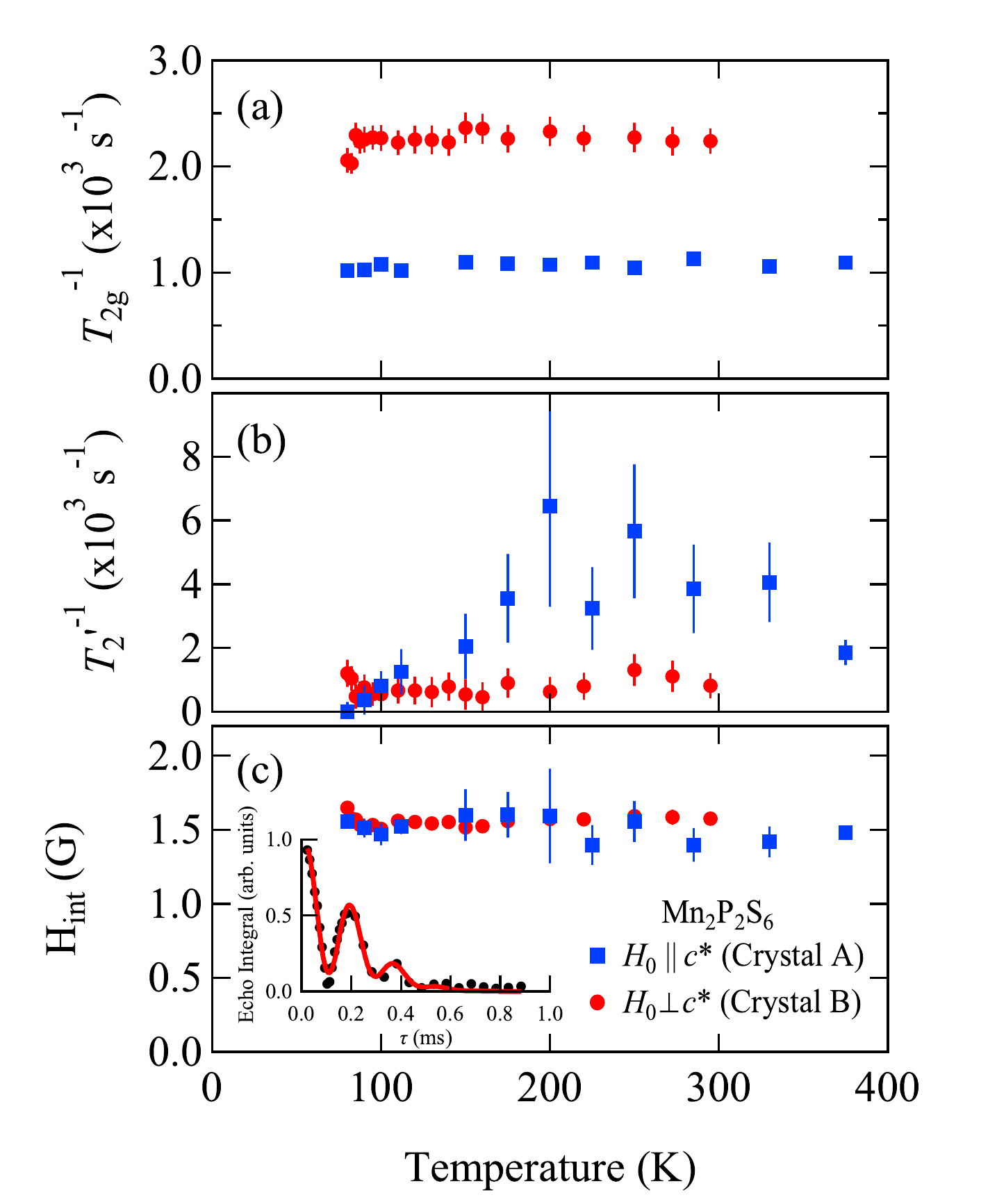}
    {\caption{\label{fig_12_Mn226_T2}Parameters $T_{2g}^{-1}$, $T_{2^\prime}^{-1}$, and 
     $H_\mathrm{int}$ extracted from fits to the echo-decay curves as a function of temperature 
     for Mn$_2$P$_2$S$_6$. The inset shows an example echo-decay curve (crystal B, 
     $H_0 \perp c^* $, $T = 295$\,K).}}
\end{figure}

We also measured the spin--spin relaxation by fitting the echo-decay curves, which show a 
combination of exponential and Gaussian behavior and an oscillatory 
behavior~\cite{Straessle_2011_AbsenceOrbitalCurrents}: 
\begin{equation}
\begin{split}
M(\tau) =& M \exp{\left[-\frac{1}{2}\left(\frac{2\tau}{T_{2g}}\right)^2\right]} \times \\
&\left(1 -F\exp{\left[-\frac{2\tau}{T^\prime_2}\right]}\cos{\left[2\tau\omega_\mathrm{int} - \psi \right]}\right),
\end{split}
\end{equation}
where $\omega_\mathrm{int} = {}^{31}\gamma H_\mathrm{int}$ and $\psi$ is a phase shift. The fit coefficients are summarized 
in Fig.~\ref{fig_12_Mn226_T2}. Gaussian-like rate component $T_{2g}$ is 
temperature independent, while the exponential rate only begins to decrease below 
approximately 200\,K for $H_0 \parallel c^*$. The extracted internal field from the oscillatory 
behavior is temperature independent.

\section{Discussion}

The most interesting result in this work is the observation of a shift anomaly in a second member 
of the $T_2$P$_2$S$_6$ family, with the first being Ni$_2$P$_2$S$_6$. Similar to the Ni case, the 
Mn sister compound has a shift anomaly that onsets approximately at $T_{\mathrm{max}}$, the 
temperature at which $\chi$ goes through a maximum. However, in the current case of 
Mn$_2$P$_2$S$_6$ the shift anomaly also shows a new behavior; not only does the $K$ fail to scale 
with $\chi$, but the value of the apparent hyperfine coupling changes sign. 

NMR shift anomalies have been observed in a few families of materials, including the 
heavy-fermion compounds CeCu$_2$Si$_2$, UPt$_3$, and 
URu$_2$Si$_2$~\cite{Shirer_2012_Longrangeorder, Curro_2004_Scalingemergentbehavior}, 
cuprates~\cite{Stemmann_1994_Spingapmagnetic, Thoma_1995_SusceptibilityKnightshift, 
Moskvin_2007_Fieldinducedstaggered}, and most recently iron-based superconductors $X$Fe$_2$As$_2$ 
($X$ = K, Rb, Cs)~\cite{Wu_2016_EmergentKondoLattice}. NMR offers a unique window into the 
emergence of electronic correlations via the NMR shift~\cite{Curro_2004_Scalingemergentbehavior}. 
More interestingly, the NMR shift anomaly has been observed to obey a universal scaling in a 
particular temperature regime across a dozen heavy-fermion materials, which has attracted 
considerable attention among various exotic behaviors of heavy-electron 
materials~\cite{Curro_2009_Nuclearmagneticresonance}.

In Ni$_2$P$_2$S$_6$ and Mn$_2$P$_2$S$_6$ there exist no conduction electrons, no structural 
transition is present that could conceivably modify the hyperfine coupling, and the lowest energy 
crystal field transition from the ground state ${}^{6}A_{1g}$ to the first excited state 
${}^{4}T_{1g}$ is 1.92\,eV~\cite{Grasso_1991_Opticalabsorptionspectra}. Therefore, another effect 
must be responsible for the $K$--$\chi$ anomaly. Considering the proximity to $T_{\mathrm{max}}$ 
in both materials, which in turn is a phenomenon associated with quasi-2D magnetic fluctuations, 
we take the observation of a shift anomaly in Mn$_2$P$_2$S$_6$ as further evidence in favor of 
this conclusion. Two other studies also find a link between quasi-2D magnetic fluctuations and 
an NMR shift anomaly~\cite{vanderKlink_2010_RelationsusceptibilityKnight, 
Sarkar_2020_Impactdisorderdynamics}. Additionally, neutron scattering directly evidences 
short-range magnetic order for $T > T_N$ in the related 
Mn$_2$P$_2$Se$_6$~\cite{Wiedenmann_1981_Neutrondiffractionstudy}.

Although we do not see any appreciable line broadening or loss of spectral weight near 
$T_{\mathrm{max}}$, it is conceivable that slow fluctuations sample off-diagonal elements of the 
hyperfine coupling tensor. Note that there is loss of spectral weight starting at about 85\,K in
Mn$_2$P$_2$S$_6$, likely due to slight broadening of the magnetic transition, though this is well 
below $T_{\mathrm{max}} \sim 117$\,K. Therefore, we calculated the dipole--dipole hyperfine coupling 
tensor of the ${}^{31}$P nuclei to the Mn magnetic moments via a lattice sum 
method~\cite{Grafe_2017_Signaturesmagneticfield}. These calculations were carried out by building 
a lattice of Mn spin sites within 600\,{\AA} radius of the ${}^{31}$P probe positions based on the
lattice parameters from Ref.~\cite{Ressouche_2010_MagnetoelectricMnPS_3as} and the fractional atomic
positions from~\cite{Ouvrard_1985_Structuraldeterminationsome}. The calculated dipolar hyperfine
coupling tensor is given by
\begin{equation}
    \label{calc_dip_hyp_tensor}
    \tilde{\mathbf{A}}_\mathrm{dip} = 
    \begin{bmatrix*}[r]
         0.033 & 0.000 &  0.001 \\
         0.000 & 0.035 &  0.000 \\
         0.001 & 0.000 & -0.067 \\
    \end{bmatrix*},
\end{equation}
where all values are given in units of T/$\mu_B$. We have rounded the values to the third decimal 
place to convey that the off-diagonal components $A_{\mathrm{dip},xz} = A_{\mathrm{dip},zx}$ 
are nonzero. We note that, to within machine precision, the other off-diagonal components are 
identically zero. The resulting calculated dipolar hyperfine fields---based on ordered
moments of Mn with magnitude 4.43 $\pm$ 0.03\,$\mu_B$ canted $\sim 8^\circ$ from $c^*$ toward the $a$ direction~\cite{Ressouche_2010_MagnetoelectricMnPS_3as}---at the two magnetically inequivalent P sites 
are $H_\mathrm{int,above}^\mathrm{dip} = -0.007\,\mathrm{T}~\hat{b}$, 
and $H_\mathrm{int,below}^\mathrm{dip} = 0.007\,\mathrm{T}~\hat{b}$. By comparing the measured 
hyperfine couplings to the values from this tensor we can estimate the contribution of transferred 
hyperfine coupling, $A_\mathrm{tr} = A_\mathrm{tot} - A_\mathrm{dip}$. These values are summarized 
in Table~\ref{transferred_hyperfine_pars_table}, and show that the diagonal components of the 
transferred hyperfine coupling tensor are of the same order of magnitude as the dipolar hyperfine 
coupling.


%

\begin{table}
    \begin{ruledtabular}
    \begin{tabular}{cccc}
              & {$A_\mathrm{tot}$ (T/$\mu_B$)} & {$A_\mathrm{dip}$ (T/$\mu_B$)} & {$A_\mathrm{tr}$ (T/$\mu_B$)} \\[0.25ex]
    \hline \\[-1.5ex]
    $A_{c^*}$ & $-0.04 \pm 0.01$               & $-0.07$                        & $ 0.05 \pm 0.01$              \\[0.5ex]
    $A_{ab}$  & $-0.02 \pm 0.01$               & $ 0.03$                        & $-0.05 \pm 0.01$              \\[0.5ex]
    \end{tabular}
    \end{ruledtabular}
    {\caption{\label{transferred_hyperfine_pars_table}Experimentally observed total hyperfine 
     coupling $A_\mathrm{tot}$, calculated dipolar coupling $A_\mathrm{dip}$, and resultant 
     transferred hyperfine coupling $A_\mathrm{tr} = A_\mathrm{tot} - A_\mathrm{dip}$. 
     $A_\mathrm{tot}$ are the high temperature values from Table~\ref{hyperfine_fit_pars_table}. 
     $A_{ab,\mathrm{dip}}$ is an average of $A_{a}$ and $A_{b}$ from 
     Eqn.~\ref{calc_dip_hyp_tensor}.}}
\end{table}

Our measurements of the internal hyperfine field in Mn$_2$P$_2$S$_6$ as a function of temperature 
are summarized in Fig.~\ref{fig_8_Mn226_Hint_vs_temp}. We have performed a multi-peak fit and then found $H_{int}$ by calculating the difference in frequency between a pair of magnetically split peaks of the form:
\begin{equation}
H_{int} = (f_2 - f_1)/(2\gamma)
\label{hint}
\end{equation}
where $\gamma$ is the ${}^{31}$P nuclear gyromagnetic ratio.

We fit the measured internal field as a 
function of temperature to a power law of the form
\begin{equation}
    H_{\mathrm{int}} (T) = H_{\mathrm{int},0}\left(1 - \frac{T}{T_{\mathrm{N}}}\right)^{\beta}
    \label{internal_field_fit_function}
\end{equation}
where $H_{\mathrm{int},0}$ is the zero temperature internal field, $T_N$ is the N\'{e}el 
temperature and $\beta$ is the power law exponent. The least squares fit results in 
$H_{\mathrm{int},0} = 0.0087 \pm 0.0001$\,T, $T_{\mathrm{N}} = 76.9 \pm 0.6$\,K, and $\beta=0.22 \pm 0.02$. 
The value of $\beta$ here agrees well with the two-dimensional anisotropic Heisenberg (2DAH) model value of $\beta = 0.231$ and the 
previously measured value of $\beta = 0.25 \pm 0.01$~\cite{Wildes_2006_Staticdynamiccritical, 
Wildes_1998_Spinwavescritical, Bramwell_1993_Magnetizationuniversalsub}. The experimentally 
determined zero temperature internal field $H_{\mathrm{int},0}$ also agrees reasonably well with 
the calculated dipolar hyperfine field. This means that the local contribution of the static
transferred hyperfine fields from the six surrounding Mn moments nearly cancel out or are individually very small.

A further magnetic state measurement that is of significant interest is the observation of 
$\pm 60^\circ$ domains in both crystals of Mn$_2$P$_2$S$_6$ that we measured, although we only show 
results from Mn$_2$P$_2$S$_6$ crystal A in Fig.~\ref{fig_7_Mn226_s1_magnetic_state_spectra}. 
Similar spectra were also observed in Ni$_2$P$_2$S$_6$ crystal B from our previous work, but 
Ni$_2$P$_2$S$_6$ crystal A was never measured. In this work we found that Ni$_2$P$_2$S$_6$ crystal 
A only has two peaks that evolve with angle in an identical fashion as the offset spectra from 
Ni$_2$P$_2$S$_6$ crystal B. This sample dependence of stacking-fault domains certainly warrants 
further investigation, ideally with surface-sensitive probes in a crystal with many step edges. 
This phenomenon may also be somewhat material dependent, as we also do not observe 
stacking-fault-induced spectral splitting in Fe$_2$P$_2$S$_6$~\cite{Peeck_2021_NMRStudyFe_2P_2S_6}.

The magnetic field dependence of the internal field also allowed us to carefully track the 
spin-flop transition in Mn$_2$P$_2$S$_6$ and resulted in the discovery of a high-field spin-flop 
transition in Ni$_2$P$_2$S$_6$. As we were limited to a maximum field of 16\,T, it would be very 
interesting to explore the evolution of this spin-flop phase in higher fields with both NMR and 
bulk probes, such as magnetization. Furthermore, a recent report of the substitution dependence 
of the spin-flop transition in 
(Ni$_{1-x}$Mn$_x$)$_2$P$_2$S$_6$~\cite{Basnet_2021_Highlysensitivespin} motivates high-field 
exploration of the substitution dependence, especially for $x<0.5$.

The relaxation data agree well with the literature for 
Mn$_2$P$_2$S$_6$~\cite{Berthier_1978_NMRinvestigationlayered}, but as no field dependence of $T_N$ 
was observed in $M$ vs $T$ at various fields up to 
7\,T~\cite{Shemerliuk_2021_TuningMagneticTransport}, we did not investigate the field dependence 
of $T_1$. Previous measurements in several $T_2$P$_2$X$_6$ compounds found field dependence of 
the magnetic transitions as well as field dependence of the spin--lattice relaxation rate. We 
speculate that the previously observed field dependence is related to sample 
quality~\cite{Berthier_1978_NMRinvestigationlayered, Ziolo_1988_31PNMRrelaxation, 
Torre_1989_Spindynamicsmagnetic}.

We found that in Mn$_2$P$_2$S$_6$ $(T_1T)^{-1}$ qualitatively tracks the bulk magnetic 
susceptibility, i.e. there is a downturn at $T_\mathrm{max}$ for both $H_{0}$ $\parallel$ $c^*$ 
and $H_{0}$ $\perp$ $c^*$. This is not uncommon in local moment insulators, and may indicate 
that fluctuations begin to become gapped even above $T_N$. However, no loss of ${}^{31}$P 
spectral weight was observed in this temperature range, which means that, to within the 
experimental uncertainty, we still observe all spins in the ensemble until approximately 85\,K.

In a local moment system at very high temperatures, we expect $T_1^{-1}$ to saturate. Considering 
the fact that we know there to be two contributions to the total hyperfine coupling, i.e. 
dipole--dipole hyperfine coupling and transferred hyperfine coupling, and the fact that dipole 
coupling is long range, whereas transferred hyperfine coupling is short range; these 
two hyperfine couplings should contribute independently to relaxation of the nuclei~\cite{Grafe_2009_ElectronicpropertiesLaO_1, Curro_2003_NuclearMagneticResonance, 
Imai_1993_Lowfrequencyspin}. We calculate the high temperature limit of $T_1^{-1}$ for 
Mn$_2$P$_2$S$_6$ based on the fluctuations of exchange-coupled moments where the total angular 
momentum can be approximated as $J \approx S$~\cite{Moriya_1956_NuclearMagneticRelaxation1, 
Moriya_1956_NuclearMagneticRelaxation2} via:
\begin{equation}
\label{T1inv_ex_hiT_lim_dip_tr}
\begin{split}
\lim_{T \rightarrow \infty}\left(\frac{1}{T_1}\right) =& \left(\left(A_\mathrm{dip}^\perp\right)^2 + \left(A_\mathrm{tr}^\perp\right)^2\right)\sqrt{2\pi}
 \left(\frac{\gamma g \mu_B}{z^\prime} \right)^2 \times\\ 
&\frac{z^\prime S(S + 1)}{3 \omega_{ex}},
\end{split}
\end{equation}
where $\gamma$ is the nuclear gyromagnetic ratio, $g$ is the electron $g$-factor, 
$A_\mathrm{dip}^\perp$ and $A_\mathrm{tr}^\perp$ are the dipolar hyperfine coupling and the 
transferred hyperfine coupling perpendicular to the applied field, respectively, $\mu_B$ is the 
Bohr magneton, $z^\prime$ is the number of coupled ${}^{31}$P sites, $S$ is the spin angular 
momentum of the magnetic ion. $\omega_{ex}$ is the Heisenberg exchange frequency given by,
\begin{equation}
\label{omega_ex}
\omega_{ex} = \frac{|J_\mathrm{ex,max}|k_B}{\hbar} \sqrt{\frac{2 z S(S + 1)}{3}},
\end{equation}
where $J_\mathrm{ex,max}$ is the maximum exchange coupling and $z$ is the number of exchange 
coupled moments~\cite{Nath_2009_Singlecrystal31P}.

The maximum of the exchange couplings is $J_1 = -0.77 \pm 0.09$\,meV, which was determined from 
neutron scattering experiments~\cite{Wildes_1998_Spinwavescritical}. The calculations for $S=5/2$ 
Mn$_2$P$_2$S$_6$ yield $\omega_{ex} = 4.9 \times 10^{12} \pm 6 \times 10^{11}\,\mathrm{s}^{-1}$, 
and the calculated dipole and experimentally extracted transferred perpendicular hyperfine 
couplings $\left|A_\mathrm{dip}^\perp(H_0 \parallel c^*)\right| = 0.03\,\mathrm{T}/\mu_B$,
$\left|A_\mathrm{tr}^\perp(H_0 \parallel c^*)\right| = 0.05 \pm 0.01\,\mathrm{T}/\mu_B$,
$\left|A_\mathrm{dip}^\perp(H_0 \perp c^*)\right| = 0.07\,\mathrm{T}/\mu_B$, and
$\left|A_\mathrm{tr}^\perp(H_0 \perp c^*)\right| = 0.05 \pm 0.01\,\mathrm{T}/\mu_B$. we obtain: 
$\lim_{T \rightarrow \infty} \left( T_1^{-1} \right)_\parallel = 40 \pm 20\,\mathrm{s}^{-1}$ 
for $H_0 \parallel c^*$ and 
$\lim_{T \rightarrow \infty}\left(T_1^{-1}\right)_\perp = 80 \pm 10\,\mathrm{s}^{-1}$ 
for $H_0 \perp c^*$. The disagreement with the experimental data (see Fig.~\ref{fig_11_Mn226_T1}) up to the highest measured temperatures may point toward the existence of
nonzero off-diagonal components of the transferred hyperfine coupling tensor.


In the case of Ni$_2$P$_2$S$_6$ we were able to perform this calculation considering only the 
total hyperfine couplings, because hyperfine coupling was dominated by the transferred component of 
the hyperfine coupling. Indeed, the high-temperature limits of $T_1^{-1}$ for Ni$_2$P$_2$S$_6$ 
based on Equation~\ref{T1inv_ex_hiT_lim_dip_tr} are slightly larger, but well within the 
uncertainties in comparison to Ref.~\cite{Dioguardi_2020_Quasitwodimensional}:
$\lim_{T \rightarrow \infty} \left( T_1^{-1} \right)_\parallel = 200 \pm 100\,\mathrm{s}^{-1}$ for 
$H_0 \parallel c^*$ and 
$\lim_{T \rightarrow \infty}\left(T_1^{-1}\right)_\perp = 140 \pm 80\,\mathrm{s}^{-1}$ 
for $H_0 \perp c^*$.
\black

Finally, we address the curious oscillatory behavior of the echo-decay curves, an example of which 
is shown in the inset of Fig.~\ref{fig_12_Mn226_T2}. These oscillations manifest as a result of 
either low energy (as compared to the scale of the linewidth) nuclear quadrupole or hyperfine 
interactions. In the case of $I = 1/2$ ${}^{31}$P, the nuclear quadrupole interaction does not come 
into play, and therefore there must be a small temperature-independent hyperfine field. This field 
$H_\mathrm{int}$ does not arise from static magnetism of Mn for $T \gg T_N$, and therefore must 
come from another source. The temperature-averaged experimental value of 
$H_\mathrm{int} = 1.53 \pm 0.02$\,G is actually quite close to the internal field expected to be 
generated by homonuclear dipole--dipole coupling between the P nuclei in the P dimers, which in the 
case of P nuclei separated by $R_\mathrm{P-P} = 2.187(3)$\,{\AA} is 
$H_\mathrm{int, Pake} = 1.639 \pm 0.007$\,G.

Based on the calculation of the splitting/angular dependence of the Pake doublet for 
Mn$_2$P$_2$S$_6$, we would expect a maximum splitting $\Delta f = 5.66\pm 0.02$\,kHz for 
$H_0 \parallel c^*$~\cite{Pake_1948_NuclearResonanceAbsorption}. We do observe a spectrum for this 
orientation with a marginally double-peaked shape, and an average temperature-independent 
splitting of $8.7 \pm 0.7$\,kHz, but the larger spectral width (average FWHM = 12.2\,kHz if fit 
with two peaks) almost completely obscures this effect. This is different from the case of 
Ni$_2$P$_2$S$_6$, where the lines were very narrow and we could observe the doublet cleanly.

\section{Conclusions}
\label{sec:conclusions}

In summary, we have conducted a detailed NMR study of Mn$_2$P$_2$S$_6$, revealing an NMR shift 
anomaly associated with quasi-2D magnetic correlations. We find this anomaly manifests in the 
vicinity of $T_\mathrm{max}$, the maximum in the static bulk magnetic susceptibility $\chi$. 
This anomaly occurs in both Mn$_2$P$_2$S$_6$ and Ni$_2$P$_2$S$_6$ in spite of a factor of two 
difference in $T_N$. In the case of Mn$_2$P$_2$S$_6$ the apparent hyperfine coupling changes sign 
in the low temperature regime of the paramagnetic state (approximately below $T_\mathrm{max}$), 
possibly revealing the existence of short-range magnetic correlations, although no appreciable 
increase in linewidth or loss of spectral weight occur. The calculated diagonal elements of the 
transferred hyperfine coupling tensor are of the same order of magnitude as those of the 
geometrical dipole--dipole hyperfine coupling tensor. $(T_1T)^{-1}$ qualitatively follows the 
peak-like behavior of $\chi$, indicating that NMR relaxation is also sensitive to quasi-2D 
correlations, which in turn points to suppression of spin fluctuations that onsets at 
$T_\mathrm{max}$.

Magnetic state NMR measurements of Mn $_2$P$_2$S$_6$ indicate that stacking faults are not 
unique to the case of Ni$_2$P$_2$S$_6$, as the spectra split into three pairs of peaks with 
$\pm 60^\circ$ rotational offset. However, upon revisting a second crystal of Ni$_2$P$_2$S$_6$, 
we observe evidence for only one magnetic domain, and therefore demonstrate sample dependence 
that motivates future measurements with techniques sensitive to the existence magnetic domains 
and/or stacking faults. We also showed that NMR is a sensitive probe of the spin-flop transition 
in Mn$_2$P$_2$S$_6$ with $H_\mathrm{sf} = 4$\,T. Field-dependent spectral measurements of 
Ni$_2$P$_2$S$_6$ in the magnetic state at high field reveal a similar spin-flop transition with 
$H_\mathrm{sf} = 14$\,T. We therefore propose further high-field investigation of this compound 
via bulk probes, especially magnetization.

\begin{acknowledgments}
The authors would like to thank P. Fritsch, N. J. Curro, and H. Yasuoka for fruitful 
discussions. A.P.D. was supported by DFG Grant No. DI2538/1-1. S.A. acknowledges financial 
support of (DFG) through Grant No AS 523/4-1. S.S. acknowledges financial support from the 
graduate academy GRK-1621 of the DFG (Project No. 129760637). A.B. and F.B. acknowledge 
financial support of Tunisian Ministry of Higher Education and Scientific Research. B.B. 
S.A. \& Y.S. acknowledge financial support of BMBF through UKRATOP (BMBF), under reference 
01DK18002.
\end{acknowledgments}


%

\end{document}